\documentclass[%
 reprint,
 superscriptaddress,
 preprintnumbers,
 nofootinbib,  
 amsmath,amssymb,
 aps,
 prl,
]{revtex4-2}

\usepackage{algorithm}
\usepackage{algpseudocode}
\usepackage{psfrag,slashed,cancel,array,graphicx,todonotes}
\usepackage{subfigure}
\usepackage[labelfont=bf,labelsep=quad,justification=justified]{caption}
\usepackage[utf8]{inputenc}
\usepackage{mathtools}
\usepackage{mathrsfs}
\usepackage{multirow}
\usepackage{braket}
\usepackage{slashed}
\usepackage{booktabs} 
\usepackage[normalem]{ulem}
\usepackage{slashed}
\usepackage{hyperref}

\hypersetup{pdftitle={A new angle on coherence-violating logarithms in color-singlet processes},pdfcreator={},linkcolor=[rgb]{0.15,0.35,0.75},colorlinks=true,citecolor=[rgb]{0.675,0,0.2},urlcolor=[rgb]{0.15,0.35,0.65}}

\newcommand{\df}{\mathrm{d}}
\newcommand{\img}{\mathrm{i}}

\newcommand{\born}{ \hat \sigma_0}

\newcommand{\T}{{\boldsymbol{T}}}
\newcommand{\G}{{\boldsymbol{\Gamma}}}
\newcommand{\D}{{\boldsymbol{D}}}
\newcommand{\g}{{\boldsymbol{\gamma}}}

\def\beq{\begin{equation}}
\def\eeq{\end{equation}}
\def\bea{\begin{eqnarray}}
\def\eea{\end{eqnarray}}

\AtBeginDocument{
\heavyrulewidth=.08em
\lightrulewidth=.05em
\cmidrulewidth=.03em
\belowrulesep=.65ex
\belowbottomsep=0pt
\aboverulesep=.4ex
\abovetopsep=0pt
\cmidrulesep=\doublerulesep
\cmidrulekern=.5em
\defaultaddspace=.5em
}

\def\be{\begin{equation}}
\def\ee{\end{equation}}

\begin{document}

\preprint{CERN-TH-2026-260}

\title{A new angle on coherence-violating logarithms in color-singlet processes}

\author{Mrinal Dasgupta}
\email{mrinal.dasgupta@manchester.ac.uk}
\affiliation{Department of Physics \& Astronomy, University of Manchester, Manchester M13 9PL, United Kingdom}

\author{Alexander Fraley}
\email{alexander.fraley@postgrad.manchester.ac.uk}
\affiliation{Department of Physics \& Astronomy, University of Manchester, Manchester M13 9PL, United Kingdom}

\author{Pier Francesco Monni}
\email{pier.monni@cern.ch}
\affiliation{CERN, Theoretical Physics Department, CH-1211 Geneva 23, Switzerland}

\author{Saad Nabeebaccus}
\email{saad.nabeebaccus@manchester.ac.uk}
\affiliation{Department of Physics \& Astronomy, University of Manchester, Manchester M13 9PL, United Kingdom}

\begin{abstract}
We introduce a new formulation for the resummation of coherence-violating logarithms for the transverse energy flow in rapidity gaps in the Drell--Yan process. Our formalism allows us to resum coherence-violating logarithms in the gap cross section in the phenomenologically important regime where the gap size is comparable to the logarithm of the transverse energy.
In the formal small gap limit, our result reduces to the known state-of-the-art resummation of coherence-violating logarithms for the Drell--Yan process. However, we identify sizable new corrections even for moderate gap widths that significantly modify the phenomenological picture for LHC applications.
This work will form an important element of future extensions to a range of global and non-global observables.

\end{abstract}

\maketitle

\paragraph*{Introduction.---}
Understanding the breakdown of strict collinear factorization in Quantum Chromodynamics (QCD) is a key challenge for precision physics at the high-luminosity LHC. This arises when space-like initial-state splittings retain quantum correlations with the rest of the hard scattering~\cite{Catani:2011st,Forshaw:2012bi}, preventing amplitudes from factorizing into universal splitting kernels and lower-multiplicity amplitudes~\cite{Schwartz:2017nmr,Buccioni:2026mfg,Chen:2026dnj,Barcaro:2026dsd}.
A well-known consequence is the appearance, in the perturbative expansion for certain observables, of logarithmic enhancements beyond those predicted by standard collinear factorization~\cite{Forshaw:2006fk,Forshaw:2008cq,Banfi:2010xy,Becher:2021zkk,Becher:2023mtx}.

A central rapidity gap in the Drell--Yan process is a particularly simple example of this phenomenon. Here, radiation above a transverse-momentum scale $Q_0$ is vetoed in an interval of width $Y$ defined with respect to the beam axis, while radiation outside remains unrestricted, as in the classic gaps-between-jets observables~\cite{Dokshitzer:1992ip,Oderda:1998en,Kidonakis:1998nf,Berger:2001ns,Appleby:2002ke,Forshaw:2009fz}. For a hard scale $Q\gg Q_0$, the gap cross section contains large logarithms $L=\ln(Q/Q_0)$ whose resummation is needed to predict the probability of passing the veto. Because the measurement covers only part of the phase space, it is sensitive to correlated soft radiation inside and outside the gap, giving rise to non-global logarithms~\cite{Dasgupta:2001sh,Dasgupta:2002bw,Banfi:2002hw}.
Importantly, it also exposes a failure of the cancellation of initial-state collinear radiation that is naively expected from the insensitivity of the measurement to radiation flying in the direction of the beam~\cite{Forshaw:2012bi,Becher:2024kmk,Becher:2025igg}.

The underlying issue is color coherence. Soft radiation at wide angles normally couples to the total color charge of a collinear system rather than resolving its constituents. Coulomb (or Glauber) gluon exchanges~\cite{Dokshitzer:2005ek,Dokshitzer:2005ig,Aybat:2008ct,Rothstein:2016bsq}, however, generate color rotations whose ordering relative to initial-state emissions matters~\cite{Angeles-Martinez:2015rna,AngelesMartinez:2016iow,Forshaw:2021rlk,Forshaw:2021fxs}. The suppression of radiation into the gap can therefore distinguish the color states produced by different emission histories, and the gap probability acquires sensitivity to radiation collinear to the incoming beams, even though that radiation lies outside the measured region. This leads to the appearance, in the perturbative expansion, of logarithmic corrections with  additional collinear enhancements, known as coherence-violating logarithms (CVLs)~\cite{Forshaw:2006fk,Forshaw:2008cq,Keates:2009dn,Forshaw:2012bi,Banfi:2025mra,Dasgupta:2025cgl,Becher:2026kbr}.

Drell--Yan production provides a simple setting in which to isolate this mechanism. At Born level the incoming quark-antiquark pair forms a color singlet, and the Coulomb contribution is an unobservable phase. A nonzero
coherence-violating correction requires radiation associated with both incoming legs, two Coulomb exchanges, and a vetoed emission. It therefore first appears at order $\alpha_s^5$. The strict double-logarithmic tower in the narrow-gap limit, $\alpha_s^k YL^{2k-3}$ for $k\geq5$, has been resummed to all orders~\cite{Becher:2021zkk,Becher:2023mtx,Boer:2023jsy,Boer:2024hzh}. This result sums arbitrarily many collinear emissions but retains only the term linear in $Y$. It therefore leaves open how the size of coherence violation changes when the gap width is not small compared with $L$.

The gap width does more than multiply the probability for an emission into the vetoed region. It also restricts the real radiation needed to make the Coulomb phases observable. In logarithmic soft kinematics, an out-of-gap emission satisfies
$Y/2<|y|<\ln(Q/k_t)$ and therefore requires $k_t<Qe^{-Y/2}$.
The available logarithmic rapidity range is consequently
$\Lambda=L-Y/2$, rather than $L$. The first nonzero term of the
resummation derived below makes the importance of this restriction explicit. In units of the Born cross section, and with
$\bar\alpha=N_c\alpha_s/\pi$, the coherence-violating correction reads
\begin{equation}
 \sigma_{\mathrm{CVL}}
 =-\frac{\pi^2}{210}\frac{N_c^2-1}{N_c^4}\,
   \bar\alpha^5Y\left(L-\frac{Y}{2}\right)^7 + {\cal O}(\bar\alpha^6).
 \label{eq:first-CVL}
\end{equation}
%

The large power accompanying the logarithmic term explains why finite-gap effects can be numerically important when comparing Eq.~\eqref{eq:first-CVL} to its narrow-gap counterpart, obtained by setting $Y=0$ in the term in brackets. One observes that already at $Y=L/10$, the finite-gap factor reduces the first nonzero contribution by about $30\%$ compared to its narrow-gap approximation. The finite-gap corrections carry higher powers of $Y$ and correspondingly lower powers of $L$ at fixed order, so they are not suppressed when $Y\sim L$.

An all-order account of finite-gap effects requires a simultaneous resummation of all terms of type $\alpha_s^k L^{2k-m-2} Y^m$ for $k\geq5$ and $1\leq m\leq 2k-2$. This requires the description of an arbitrary number of soft emissions inside the gap. Their effect depends on the distribution of radiation outside the gap, and its color representation. The required formulation of the resummation must simultaneously retain the restricted
collinear phase space and the color-dependent soft evolution, and cannot be obtained merely by shifting $L$ in the strict narrow-gap result.
This Letter presents a new formulation of the resummation of coherence-violating logarithms, and showcases it by resumming, for the first time, the leading tower of CVLs in the finite-gap case, while retaining exact information about the number of colors $N_c$. Remarkably, the answer is encoded in a simple integral representation for the coherence-violating correction, that can be evaluated with arbitrary precision.
The impact of our calculation on the phenomenology of CVLs is significant, in that we observe an important reduction of the CVL corrections compared to the narrow-gap approximation, highlighting the importance of incorporating finite-gap effects in the description of coherence-violating logarithms.

\begin{figure}[t]
\includegraphics[width=\columnwidth]{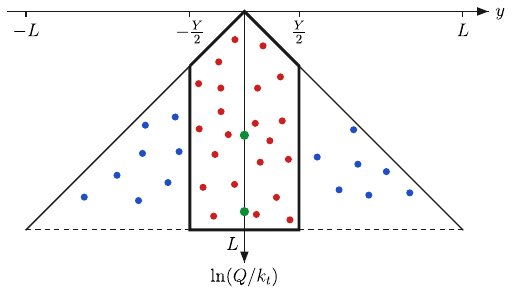}
\caption{The rapidity-gap phase space on the Lund plane. Blue: out-of-gap emissions; red: in-gap emissions, vetoed above $Q_0$; green: the two Coulomb exchanges.}
\label{fig:lund}
\end{figure}

\paragraph*{The master equation.---}
We consider an incoming quark $a$ and antiquark $b$, which produce a color-singlet final state at hard scale $Q$, with radiation into rapidities $|y|<Y/2$ vetoed above $Q_0\ll Q$. The cross section is expressible as a sum over $n$ soft gluons emitted outside the gap, ordered in their transverse momentum $k_t$ with respect to the $ab$ dipole,
\begin{align}
\sigma &= \sum_{n}\bar\alpha^{\,n}\!\int_{\mathrm{out}}\!\prod_{l=1}^{n}\df\Phi_l\,\big\langle f\big|\Big(\mathcal{U}(k_{t1},Q)\circ\hat{\mathcal{P}}^{(1)}\circ\mathcal{U}(k_{t2},k_{t1}) \nonumber\\
&\qquad\circ\cdots\circ\hat{\mathcal{P}}^{(n)}\circ\mathcal{U}(Q_0,k_{tn})\Big)[\boldsymbol{1}]\big|f\big\rangle\,, \label{eq:master}
\end{align}
with $\df\Phi_l=(\df k_{tl}/k_{tl})\,\df y_l\,\df\phi_l/2\pi$ the phase space of emission $l$, with transverse momentum $k_{tl}$, rapidity $y_l$ and azimuth $\phi_l$, ordered as $Q_0<k_{tn}<\dots<k_{t1}<Q$. 
With $R(k_t)=\min\big(Y/2,\ln(Q/k_t)\big)$ the boundary of the veto region, the subscript ``out'' restricts each emission to $R(k_{tl})<|y_l|<\ln(Q/k_{tl})$, see Fig.~\ref{fig:lund}. The ket $|f\rangle$ is the Born color state, normalized so that $\sigma$ is in units of the Born cross section $\born$.

Eq.~\eqref{eq:master} is built outward from the identity $\boldsymbol{1}$ at $Q_0$ by two maps on color operators, $\hat{\mathcal{P}}^{(l)}$ and $\mathcal{U}$. The map $\hat{\mathcal{P}}^{(l)}$ adds emission $l$, real plus virtual. In the color-space notation of~\cite{Catani:1996vz},
\begin{align}
\hat{\mathcal{P}}^{(l)}[\boldsymbol{X}] &= -\D^{(l)\mu\dagger}\boldsymbol{X}_{\mathrm{r}}\,\D^{(l)}_{\mu}-\{\g,\,\boldsymbol{X}_{\mathrm{v}}\}, \nonumber\\
\D^{(l)\mu} &= \sum_i\T_i^{(l)} h_i^\mu\,, \qquad \g=-\tfrac12\sum_{i<j}\T_i\!\cdot\!\T_j\,\omega_{ij}\,, \nonumber\\
h_i^\mu &= \frac{k_{tl}}{2}\frac{p_i^\mu}{p_i\cdot k_l}\,, \qquad \omega_{ij}=2\,h_i\cdot h_j=\frac{k_{tl}^2}{\big(k_{tl}^{(ij)}\big)^2}\,, \label{eq:current}
\end{align}
where $\boldsymbol{X}$ is the color operator of the softer emissions, written $\boldsymbol{X}_{\mathrm{r}}$ with gluon $l$ present and $\boldsymbol{X}_{\mathrm{v}}$ without. The sums in the real emission matrix $\D^{(l)}$ and virtual counterpart $\g$ run over the legs $i$, $j$ that can radiate gluon $l$: $a$, $b$ and the real gluons $1,\dots,l-1$. $\T_i^{(l)}$ corresponds to the emission of gluon $l$ from leg $i$, of momentum $p_i$. Gluon $l$ has momentum $k_l$, and $\big(k_{tl}^{(ij)}\big)^2=2\,(k_l\cdot p_i)(k_l\cdot p_j)/(p_i\cdot p_j)$ is its transverse momentum with respect to the $(ij)$ dipole.

The map $\mathcal{U}$ evolves the color operator between emissions, $\mathcal{U}(k,k')[\boldsymbol{X}]=\boldsymbol{U}^\dagger(k,k')\,\boldsymbol{X}\,\boldsymbol{U}(k,k')$, with
\begin{align}
\boldsymbol{U}(k,k') &= \mathrm{P}\exp\Big[-\bar\alpha\!\int_k^{k'}\!\frac{\df k_t}{k_t}\,\G(k_t)\Big]\,, \nonumber\\
\G(k_t) &= \hat{\G}_R(k_t)+\img\pi\,\T_a\!\cdot\!\T_b\,, \nonumber\\
\hat{\G}_R(k_t) &= -\tfrac12\sum_{i<j}\T_i\!\cdot\!\T_j\int_{\mathrm{in}}\!\frac{\df y\,\df\phi}{2\pi}\,\omega_{ij}\,, \label{eq:evolution}
\end{align}
where $\mathrm{P}$ orders lower $k_t$ to the left. The sum runs over $a$, $b$ and the real gluons harder than $k_t$, and $\omega_{ij}$ in Eq.~\eqref{eq:current} corresponds to  a vetoed gluon of transverse momentum $k_t$, with its rapidity restricted to the gap (``in''). Using the above definitions of $\mathcal{U}$ and $\hat{\mathcal{P}}$, Eq.~\eqref{eq:master} takes its usual nested structure.

A key aspect of Eq.~\eqref{eq:master} is its kinematic ordering: the soft emissions, real and virtual, and the Coulomb gluon exchanges are all ordered in transverse momentum with respect to the $ab$ dipole.
Refs.~\cite{Angeles-Martinez:2015rna,AngelesMartinez:2016iow,Forshaw:2021rlk} showed, at fixed perturbative order, that the soft emissions and the Coulomb exchanges must be ordered in their transverse momentum with respect to the emitting dipoles. In the leading-logarithmic approximation (strong ordering in both $k_t$ and rapidity), this coincides with the ordering with respect to the $ab$ dipole.
%

%

At $n=0$, the imaginary part of $\G$, the Coulomb exchange $\img\pi\,\T_a\cdot\T_b$, becomes a pure phase and cancels, and Eq.~\eqref{eq:master} gives the usual Sudakov suppression,
\begin{equation}
\sigma\big|_{n=0}=\big\langle f\big|\boldsymbol{U}^\dagger(Q_0,Q)\,\boldsymbol{U}(Q_0,Q)\big|f\big\rangle=e^{-\bar\alpha Y(L-Y/4)\lambda_f}\,, \label{eq:sudakov}
\end{equation}
with $L=\ln(Q/Q_0)$ and $\lambda_f=\T_a^2=(N_c^2-1)/N_c^2$ for color charges in units of $\sqrt{N_c/2}$. The exponent contains $Y(L-Y/4)$, the area of the veto region in Fig.~\ref{fig:lund}. Without the Coulomb exchange, the terms with $n\ge1$ give the non-global logarithms~\cite{Dasgupta:2001sh,Dasgupta:2002bw,Banfi:2002hw}. As we will see, the presence of the Coulomb exchange generates CVLs~\cite{Forshaw:2006fk,Forshaw:2008cq}, from $n\ge2$ for the Drell--Yan process.

We resum the leading CVLs, which at order $\bar\alpha^k$ carry $2k-2$ powers of $L$ and $Y$, counting $Y\sim L$, as in Eq.~\eqref{eq:first-CVL}. Every soft gluon, real or vetoed, must then give two logarithms and each Coulomb exchange one. We therefore evaluate Eq.~\eqref{eq:master} in the soft-collinear limit, where the emissions, already strongly ordered in $k_t$, are also widely separated in rapidity. This allows a key simplification: $\omega_{ij}\to\Theta_{ij}$, with $\Theta_{ij}=1$ when the emission lies between legs $i$ and $j$ in rapidity and $0$ otherwise. The error this introduces lies two logarithms below the leading CVLs~\cite{SM}. Only the dipoles spanning the gap then survive in the evolution, and color conservation sums them to one charge,
\begin{equation}
\hat{\G}_R(k_t)\to\G_R(k_t)=R(k_t)\,\T_R^2\,,\;\; \T_R={\textstyle\sum_{y_i>0}}\T_i\,, \label{eq:GRgap}
\end{equation}
with $\T_R$ the total color on leg $a$'s side of the gap, taken at $y>0$, gluons included. In this limit each out-of-gap gluon is collinear to $a$ or $b$ and, by color coherence, is emitted by the total charge of the partons more forward, i.e.\ higher $|y|$, than it, which color conservation reduces to the charge of that incoming leg. The real--virtual combination $\hat{\mathcal{P}}^{(l)}$ thus reduces to one pair per leg,
\begin{gather}
\int_{\mathrm{out}}\!\frac{\df y_l\,\df\phi_l}{2\pi}\,\hat{\mathcal{P}}^{(l)}\;\to\;v_l\,\mathcal{P}^{(l)}\,,\;\; \mathcal{P}^{(l)}=\mathcal{P}_a^{(l)}+\mathcal{P}_b^{(l)}\,, \nonumber\\
\mathcal{P}_i^{(l)}[\boldsymbol{X}]=\T_i^{(l)\dagger}\boldsymbol{X}_{\mathrm{r}}\,\T_i^{(l)}-\tfrac12\big(\T_i^2\boldsymbol{X}_{\mathrm{v}}+\boldsymbol{X}_{\mathrm{v}}\T_i^2\big)\,, \label{eq:pair}
\end{gather}
with $v_l=\ln(Q/k_{tl})-Y/2$ the width of the out-of-gap region, which closes at $k_{tl}=Qe^{-Y/2}$. Eq.~\eqref{eq:master} becomes
\begin{align}
\sigma &= \sum_n\bar\alpha^{\,n}\!\int_{Q_0}^{Qe^{-Y/2}}\!\frac{\df k_{tn}}{k_{tn}}\,v_n\cdots\!\int_{k_{t2}}^{Qe^{-Y/2}}\!\frac{\df k_{t1}}{k_{t1}}\,v_1 \nonumber\\
&\quad\times\big\langle f\big|\,\Big(\mathcal{U}(k_{t1},Q)\circ\mathcal{P}^{(1)}\circ\mathcal{U}(k_{t2},k_{t1}) \nonumber\\
&\qquad\circ\cdots\circ\mathcal{P}^{(n)}\circ\mathcal{U}(Q_0,k_{tn})\Big)[\boldsymbol{1}]\,\big|f\big\rangle\,, \label{eq:target}
\end{align}
with $\mathcal{U}$ now evolved by $\G_R$ in place of $\hat{\G}_R$. Through the replacement $\omega_{ij}\to\Theta_{ij}$, Eq.~\eqref{eq:target} is insensitive to non-global logarithms, which originate from wide-angle soft dynamics sensitive to the difference $\omega_{ij}-\Theta_{ij}$. These start at $\bar\alpha^2L^2$, but dropping them is justified for the leading CVLs~\cite{SM}.

\paragraph*{The two-Coulomb expansion.---}
The terms of Eq.~\eqref{eq:target} with $n\ge1$ are nonzero only because of the Coulomb phase in $\boldsymbol{U}$: $\T_R^2$ is unchanged by emissions, so the real and virtual parts of each $\mathcal{P}_i^{(l)}$ in Eq.~\eqref{eq:pair} cancel unless a Coulomb exchange stands between them. We therefore expand every $\boldsymbol{U}$ in Eq.~\eqref{eq:target} in the number of Coulomb exchanges,
\begin{gather}
\boldsymbol{U}(k,k')=\boldsymbol{S}_0+\boldsymbol{S}_1+\boldsymbol{S}_2+\dots\,,\quad \boldsymbol{S}_0=\boldsymbol{V}(k,k')\,, \nonumber\\
\boldsymbol{S}_1=-\img\pi\bar\alpha\!\int_k^{k'}\!\frac{\df\kappa}{\kappa}\,\boldsymbol{V}(k,\kappa)\,\T_a\cdot\T_b\,\boldsymbol{V}(\kappa,k')\,, \label{eq:Sexpand}
\end{gather}
with $\boldsymbol{V}(k,k')=\exp[-\bar\alpha\int_k^{k'}(\df k_t/k_t)\,\G_R(k_t)]$ the Coulomb-free evolution and $\boldsymbol{S}_2$ the same with two exchanges. Without an exchange, every term with $n\ge1$ cancels, leaving the Sudakov factor~\eqref{eq:sudakov}. Terms with one exchange vanish when their color operator is evaluated in the Born state $|f\rangle$~\cite{Forshaw:2008cq,Forshaw:2012bi}. The leading CVLs come from the terms with two, as each further pair costs two logarithms. One of the two must lie below every emission, in $\mathcal{U}(Q_0,k_{tn})$: otherwise $\mathcal{P}^{(n)}$ acts on a function of $\T_R^2$ alone, which by Eq.~\eqref{eq:pair} it maps to zero, in line with~\cite{Forshaw:2006fk,Forshaw:2008cq}. Outside the two exchanges, $\T_R^2$ keeps its Born value $\lambda_f$ since emissions leave it unchanged, and every $\boldsymbol{V}$ gives only a scalar Sudakov-like exponential. Between them, the first exchange has rotated the color and $\T_R^2$ is no longer fixed. We therefore insert $\boldsymbol{1}=\sum_\rho\boldsymbol{P}_\rho$ between the exchanges, with $\boldsymbol{P}_\rho$ the projector onto the eigenvalue $\lambda_\rho$ of $\T_R^2$. In each channel $\rho$ the $\boldsymbol{V}$ between the exchanges are then numbers too, with exponents set by $\lambda_\rho$, the squared color charge on leg $a$'s side of the gap. Collecting the terms with the exchanges at fixed $w<w'$, where $w=\ln(Q/\kappa)-Y/2$, the two-Coulomb part of Eq.~\eqref{eq:target}, which we denote $\sigma_{\mathrm{CVL}}$, becomes
\begin{align}
\sigma_{\mathrm{CVL}} &= -2\pi^2\bar\alpha^2\,e^{-\bar\alpha Y(L-Y/4)\lambda_f}\sum_{n\ge1}\bar\alpha^{\,n}\!\int\![\df v]_n\sum_{\rho\neq f} \nonumber\\
&\times\int_{v_n}^{\Lambda}\!\df w'\sum_{j=1}^{n}\int_{v_j}^{v_{j+1}}\!\df w\;\Phi_\rho(w,w')\,\Delta^{(f\rho f)}(n,j)\,, \label{eq:masterreorg}
\end{align}
where the emission integrals of Eq.~\eqref{eq:target} are $\int[\df v]_n=\int_{0<v_1<\dots<v_n<\Lambda}\prod_l v_l\,\df v_l$, with $\Lambda=L-Y/2$ and $v_{n+1}\equiv w'$. The exponential is the Sudakov factor~\eqref{eq:sudakov}. The weight $\Phi_\rho$ measures the modified suppression of the gap while the color is rotated into channel $\rho$, summed over the four placements of the two exchanges across the cut (i.e.~whether they are placed in the amplitude or complex conjugate amplitude),
\begin{equation}
\Phi_\rho(w,w')=e^{-\beta\delta_\rho(w'-w)}-e^{-\beta\delta_\rho(2\Lambda-w-w')}\,, \label{eq:Phidef}
\end{equation}
with $\beta=\bar\alpha Y$ and $\delta_\rho=(\lambda_\rho-\lambda_f)/2$. At $\rho=f$, $\delta_f=0$ and the two terms cancel, which removes the diagonal channel from Eq.~\eqref{eq:masterreorg}. The softer exchange lies at $w'$, below every emission, and the harder between emissions $j$ and $j+1$. All placements share the color factor
\begin{align}
\Delta^{(f\rho f)}(n,j) &= \big\langle f\big|\big(\mathcal{P}^{(1)}\circ\cdots\circ\mathcal{P}^{(j)}\circ\mathcal{C}\circ\Pi_\rho \nonumber\\
&\quad\circ\mathcal{P}^{(j+1)}\circ\cdots\circ\mathcal{P}^{(n)}\circ\mathcal{C}\big)[\boldsymbol{1}]\big|f\big\rangle\,, \label{eq:Delta2}
\end{align}
with $\mathcal{C}[\boldsymbol{X}]=\T_a\!\cdot\!\T_b\,\boldsymbol{X}$ the Coulomb exchange and $\Pi_\rho[\boldsymbol{X}]=\boldsymbol{P}_\rho\boldsymbol{X}$ the projection.

The Coulomb exchange mixes the charge on leg $a$'s side of the gap into $\mathrm{fund}\otimes\mathrm{adj}=\mathrm{fund}\oplus R_+\oplus R_-$. Hence $\rho\in\{f,\pm\}$, with $\lambda_\pm=\lambda_f+2\pm2/N_c$, and only $\rho=\pm$ enter Eq.~\eqref{eq:masterreorg}. For $\rho\neq f$ and $1\leq j\leq n$, we find~\cite{SM}
\begin{equation}
\Delta^{(f\rho f)}(n,j)=(-1)^n\,2^{n-j}\big(2^j-2\big)\,\frac{\dim R_\rho}{N_c^5}\,, \label{eq:Deltaclosed}
\end{equation}
with $\dim R_\pm=N_c(N_c\mp1)(N_c\pm2)/2$.

\begin{figure}[t]
\includegraphics[width=\columnwidth]{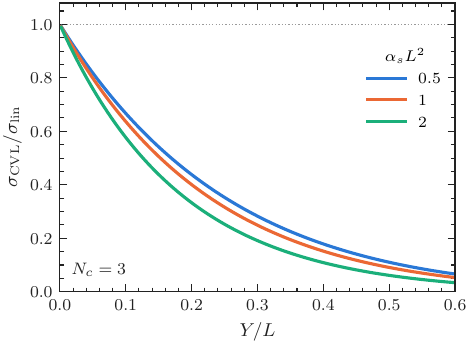}
\caption{Ratio of Eq.~\eqref{eq:resummed} to its term linear in $Y$, Eq.~\eqref{eq:linearY}, against $Y/L$, at fixed $\alpha_s L^2$ and $N_c=3$.}
\label{fig:convergence}
\end{figure}

\begin{figure}[t]
\includegraphics[width=\columnwidth]{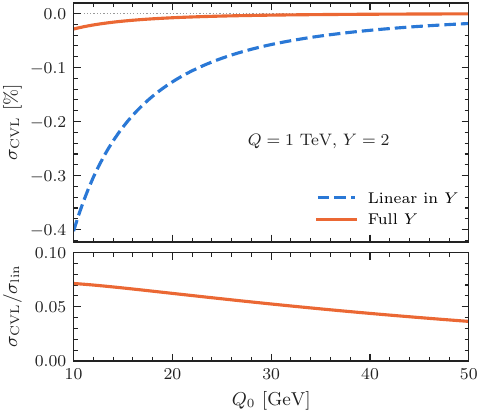}
\caption{The resummed coherence-violating correction $\sigma_{\mathrm{CVL}}$ at the kinematics of Fig.~11 of~\cite{Becher:2023mtx}: $Q=1$\,TeV, $Y=2$, with $\bar\alpha=N_c\alpha_s(\bar\mu)/\pi$ at $\bar\mu=\sqrt{QQ_0}$, two-loop running from $\alpha_s(m_Z)=0.118$ and the two-loop cusp correction. The dashed curve is the term linear in $Y$, Eq.~\eqref{eq:linearY}; the lower panel is the ratio of the full result to it.}
\label{fig:fig11}
\end{figure}

\paragraph*{The resummation.---}
Inserting Eq.~\eqref{eq:Deltaclosed} into Eq.~\eqref{eq:masterreorg} and performing the $v$ integrals and the sums over $n$ and $j$ gives
\begin{align}
\sigma_{\mathrm{CVL}} &= -2\pi^2\bar\alpha^2\,e^{-\bar\alpha Y(L-Y/4)\lambda_f}\sum_{\rho=\pm}\frac{\dim R_\rho}{N_c^5}\int_0^{\Lambda}\!\df w' \nonumber\\
&\quad\times\int_0^{w'}\!\df w\,\big(1-e^{-\bar\alpha w^2/2}\big)^2\,e^{-\bar\alpha(w'^2-w^2)}\,\Phi_\rho(w,w')\,. \label{eq:resummed}
\end{align}
Equation~\eqref{eq:resummed} resums the leading CVLs at exact $N_c$ and for any $Y/L\leq2$. Its lowest order reproduces Eq.~\eqref{eq:first-CVL}. The gap width enters through the range $\Lambda$, the Sudakov factor and $\Phi_\rho$.

In order to assess the impact of the resummation derived in this Letter, it is instructive to compare Eq.~\eqref{eq:resummed} to its small-$Y$ expansion, where we only retain the linear term in the gap width $Y$, $\sigma_{\rm lin}$. At fixed $\bar\alpha L^2$, this is obtained via the replacements $\Lambda\to L$, $e^{-\bar\alpha Y(L-Y/4)\lambda_f}\to1$ and $\Phi_\rho\to2\beta\delta_\rho(L-w')$. With $w=s\,w'$ the $w'$ integral is elementary, and yields
\begin{equation}
\sigma_{\rm lin}=-\frac{2\pi^2}{3}\,Y\,\frac{N_c^2-1}{N_c^4}\,\bar\alpha^3L^3\Big[\Sigma(0)-2\Sigma\big(\tfrac12\big)+\Sigma(1)\Big]\,, \label{eq:linearY}
\end{equation}
where $\Sigma(v)=\frac32\int_0^1\df s\,\big[2/y^2-\sqrt\pi\,\mathrm{erf}(y)/y^3\big]$ with $y^2=\bar\alpha L^2\big(1-(1-v)s^2\big)$. 
This result agrees with the state of the art resummation of CVLs derived in Ref.~\cite{Becher:2023mtx} (cf.~their Eq.~(7.7)). In Figs.~\ref{fig:convergence} and~\ref{fig:fig11}, we compare the two calculations. As $Y/L\to0$ the full result~\eqref{eq:resummed} approaches $\sigma_{\rm lin}$, but it falls away quickly at finite $Y/L$. As a case study, it is convenient to consider the kinematics of Fig.~11 of~\cite{Becher:2023mtx}, where $Y/L=0.43$--$0.67$. As shown in Fig.~\ref{fig:fig11}, Eq.~\eqref{eq:resummed} is $4$--$7\%$ of the $Y$-linear prediction, highlighting the critical role of finite-width effects for phenomenology also in regimes with relatively narrow rapidity gaps.

\paragraph*{Outlook.---}
We have developed a new formulation of CVL resummation in the rapidity-gap problem that treats the color evolution of soft radiation while retaining the full dependence on the gap geometry. Applied to Drell--Yan production, it yields a compact, exact-$N_c$ result for the leading tower of CVLs arising from the exchange of two Coulomb gluons and an arbitrary number of soft emissions, that encodes the correct description of the coherence-violating correction for any hierarchy between the gap width $Y$ and the veto logarithm $L$.

Phenomenologically, the substantial suppression relative to the narrow-gap result in the kinematics studied here demonstrates that logarithmic enhancement alone does not determine the size of CVLs. Retaining the finite rapidity range is essential to assess their impact reliably. Our result therefore provides both an analytic benchmark for future CVL calculations in Monte Carlo algorithms~\cite{AngelesMartinez:2018cfz,Forshaw:2019ver,Nagy:2019bsj,Nagy:2019pjp,DeAngelis:2020rvq,Forshaw:2025bmo,Schalch:2026xxk}, and a step towards quantifying effects beyond strict collinear factorization in the high-luminosity LHC precision program.
We stress, however, that the calculation of subleading logarithmic corrections will be paramount for an accurate phenomenological study of CVLs at colliders, building on recent progress in the understanding of CVLs~\cite{Forshaw:2021fxs,Boer:2023jsy,Boer:2023ljq,Boer:2024hzh,Boer:2024xzy,Dasgupta:2025cgl,Banfi:2025mra,Becher:2026kbr,Becher:2026uih} and non-global~\cite{Becher:2015hka,Becher:2016mmh,Schwartz:2014wha,Khelifa-Kerfa:2015mma,Khelifa-Kerfa:2024udm,Hatta:2013iba,Hagiwara:2015bia,Hatta:2020wre,DeAngelis:2020rvq,Caron-Huot:2015bja,Banfi:2021owj,Banfi:2021xzn,Becher:2021urs,Becher:2023vrh,FerrarioRavasio:2023kyg} logarithms.

The color-space organization of the formalism also provides a route to more complex jet-production processes~\cite{Forshaw:2008cq,Keates:2009dn,DuranDelgado:2011tp,Becher:2024nqc}, where additional hard colored partons enlarge the space of color correlations~\cite{Becher:2023mtx,Boer:2023ljq,Banerjee:2025kkq,Kyrieleis:2005dt}.
Furthermore, a particularly important direction is the application to CVL resummation in global collider observables \cite{Banfi:2010xy,Gaunt:2014ska,Zeng:2015iba,Forshaw:2021fxs,Banfi:2025mra,Becher:2026kbr} and in vetoes on the transverse-energy flow in an azimuthal region~\cite{Dasgupta:2025cgl}, for which the rapidity range available to soft radiation is parametrically comparable to the logarithm $L$ of the observable.

The above generalizations of the results presented here will be instrumental in shedding light on the broader description of coherence violation at colliders.

\paragraph*{Acknowledgements.---}
We thank Jack Helliwell for useful discussions on the topic of CVLs in gap cross sections.  
We thank the U.K.'s Science and Technology Facilities Council for funding the project via grants ST/X00077X/1 (MD and SN) and ST/C004233/1 (MD), and via a Ph.D. studentship (AF).
%
The work of SN was also supported by the Royal Society through Grant No. URF/R1/201500.
The work of PM is funded by the European Union (ERC, grant agreement No. 101044599). Views and opinions expressed are however those of the authors only and do not necessarily reflect those of the European Union or the European Research Council Executive Agency. Neither the European Union nor the granting authority can be held responsible for them.

\bibliographystyle{apsrev4-1}
\bibliography{refs}

\clearpage
\onecolumngrid
\raggedbottom
\appendix


\setcounter{equation}{0}
\counterwithout{equation}{section}
\renewcommand{\theequation}{S\arabic{equation}}
\setcounter{figure}{0}
\renewcommand{\thefigure}{S\arabic{figure}}

\makeatletter
\renewcommand\@biblabel[1]{[#1S]}
\makeatother

\setcounter{secnumdepth}{3}

\section*{Supplemental material}
We report some technical details needed to reproduce the results of the Letter.

\section{The replacement \texorpdfstring{$\omega_{ij}\to\Theta_{ij}$}{omega -> Theta}}\label{app:theta}

We write
\begin{equation}
\omega_{ij}=\Theta_{ij}+\delta\omega_{ij}\,, \label{eq:split}
\end{equation}
and let $\delta\sigma_{\mathrm{CVL}}^{(n)}$ be the change in the two-Coulomb part of Eq.~\eqref{eq:master} at $\bar\alpha^n$ away from the pure $\Theta_{ij}$ approximation, which gives the leading CVLs $\sigma_{\mathrm{CVL}}^{(n)}$. $\sigma_{\mathrm{CVL}}^{(n)}$ carries $2n-2$ powers of $L$ and $Y$, and we show that $\delta\sigma_{\mathrm{CVL}}^{(n)}$ carries $2n-4$.

A remainder $\delta\omega_{ij}$ can enter in two places. It can enter in the gap, in the integral of the vetoed virtual gluon over the gap in Eq.~\eqref{eq:evolution}. It can also enter at a real emission, in the weight of the out-of-gap gluon in Eq.~\eqref{eq:current}. We call these a gap remainder and an emission remainder, and group the terms of $\delta\sigma_{\mathrm{CVL}}^{(n)}$ by the number and kind of remainders they contain: G, exactly one gap remainder; E, exactly one emission remainder; GG, two or more gap remainders and no emission remainder; EE, two or more emission remainders and no gap remainder; GE, at least one of each. So
\begin{equation}
\delta\sigma_{\mathrm{CVL}}^{(n)}=\delta\sigma^{(n)}_{\mathrm{G}}+\delta\sigma^{(n)}_{\mathrm{E}}+\delta\sigma^{(n)}_{\mathrm{GG}}+\delta\sigma^{(n)}_{\mathrm{GE}}+\delta\sigma^{(n)}_{\mathrm{EE}}\,. \label{eq:classes}
\end{equation}

\begin{figure}[!t]
\centering
\includegraphics[width=0.5\textwidth]{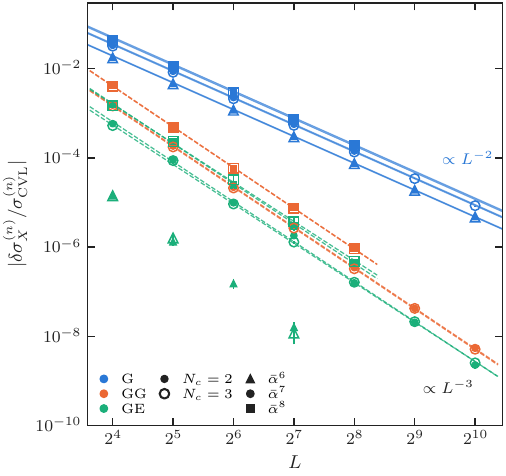}
\caption{The magnitudes of the classes of $\delta\sigma_{\mathrm{CVL}}^{(n)}$, Eq.~\eqref{eq:classes}, relative to the leading CVLs $\sigma_{\mathrm{CVL}}^{(n)}$ at the same order, against $L$ at $Y/L=1/2$, for $N_c=2$ (filled) and 3 (open) at $\bar\alpha^6$ (triangles), $\bar\alpha^7$ (circles) and $\bar\alpha^8$ (squares). Points are the numerical evaluation, with statistical errors; points within two standard errors of zero are omitted. G and GG are positive, GE is negative at $\bar\alpha^7$ and $\bar\alpha^8$ and positive at $\bar\alpha^6$, where GG vanishes. Solid lines are the leading G term, Eq.~\eqref{eq:Glead}; they are not fits. Dashed lines are fits of $A\,L^p$ to GG and GE at $\bar\alpha^7$ and $\bar\alpha^8$, with $p$ between $-3.04$ and $-2.91$. E (below $3\times10^{-9}$) and EE (below $2\times10^{-7}$) are not shown.}
\label{fig:classes}
\end{figure}

At $\bar\alpha^5$, the lowest order of the CVLs, $\delta\sigma_{\mathrm{CVL}}^{(5)}=0$. We evaluate each class numerically at $\bar\alpha^6$, $\bar\alpha^7$ and $\bar\alpha^8$, at $N_c=2$ and 3 and at $Y/L=1/4$ and $1/2$. Fig.~\ref{fig:classes} shows the results at $Y/L=1/2$, and the pattern is clear from it. Relative to $\sigma_{\mathrm{CVL}}^{(n)}$, G falls as $L^{-2}$, so it lies two logarithms below the leading CVLs, while GG and GE fall as $L^{-3}$ and lie three logarithms below. E and EE are negligible. The same holds at $Y/L=1/4$. The leading term of $\delta\sigma_{\mathrm{CVL}}^{(n)}$ is therefore that of $\delta\sigma^{(n)}_{\mathrm{G}}$.

The leading term, $\delta\sigma^{(n)}_{\mathrm{G}}$, can be understood analytically. Its single remainder sits in the integral of the vetoed virtual gluon over the gap,
\begin{equation}
\delta\Omega_{ij}=\int_{\mathrm{in}}\!\frac{\df y\,\df\phi}{2\pi}\,\delta\omega_{ij}\,, \label{eq:dOmegadef}
\end{equation}
which corrects the gap integral of $\Theta_{ij}$, the source of the factor $Y$ for each dipole spanning the gap. $\delta\omega_{ij}$ is non-negligible only where the in-gap gluon lies within $O(1)$ in rapidity of $i$ or $j$. For a real gluon $i$ outside the gap, at distance $u_i$ in rapidity from the edge, this needs the in-gap gluon near the same edge. Averaged over the azimuth, which is exact in $\delta\sigma^{(n)}_{\mathrm{G}}$ because every other weight is $\Theta_{ij}$, this gives~\cite{Kyrieleis:2005dt,Forshaw:2006fk,Forshaw:2008cq}
\begin{equation}
\delta\Omega_{ij}=\xi(u_i)+\xi(u_j)\,,\qquad \xi(u)=\frac12\ln\frac{1-e^{-2u-2Y}}{1-e^{-2u}}\,, \label{eq:xi}
\end{equation}
with $\xi=0$ for the incoming legs. $\delta\Omega_{ij}$ carries no factor $Y$: one logarithm is lost. In Eq.~\eqref{eq:evolution} each $\delta\Omega_{ij}$ comes with the color factor of its dipole, $-\frac12\T_i\cdot\T_j$. On the color singlet $\sum_{j\neq i}\T_i\cdot\T_j=-\T_i^2$. With color charges in units of $\sqrt{N_c/2}$, as in the main text, $\T_g^2=2$ for a gluon, so the sum over dipoles collapses to a number,
\begin{equation}
-\frac12\sum_{i<j}\T_i\cdot\T_j\,\delta\Omega_{ij}=\sum_g\xi(u_g)\,, \label{eq:cnum}
\end{equation}
with $g$ running over the real gluons present, which lie outside the gap. Since $\xi(u)$ falls as $e^{-2u}$ away from the edge, the remainder pins gluon $g$ close to the edge of the gap: its rapidity integral, which with $\Theta_{ij}$ gives the factor $v_g$ in Eq.~\eqref{eq:target}, now has weight only there and becomes
\begin{equation}
\int_0^\infty\!\df u\;\xi(u)=\frac{\zeta_2-\mathrm{Li}_2(e^{-2Y})}{4}=\frac{\zeta_2}{4}+\mathcal{O}\big(e^{-2Y}\big)\,, \label{eq:intxi}
\end{equation}
and a second logarithm is lost. Relative to $\sigma_{\mathrm{CVL}}^{(n)}$, then, one insertion of $\hat{\G}_R$ has its gap integrals replaced by $\delta\Omega_{ij}$, which by Eq.~\eqref{eq:cnum} gives a number with no factor $Y$, and the gluon $g$ it pins has its rapidity integral $v_g$ replaced by $\zeta_2/4$. Hence
\begin{equation}
\delta\sigma^{(n)}_{\mathrm{G}}=\frac{\zeta_2}{4}\,c_n(Y/L)\,\bar\alpha^nL^{2n-4}+\mathcal{O}\big(\bar\alpha^nL^{2n-5}\big)\,, \label{eq:Glead}
\end{equation}
where $c_n(Y/L)\,\bar\alpha^nL^{2n-4}$ is the two-Coulomb part of Eq.~\eqref{eq:target} with these two replacements, summed over $g$ and its side of the gap. Its ratio to $\sigma_{\mathrm{CVL}}^{(n)}$ falls as $L^{-2}$ at fixed $Y/L$, and is the line for $\delta\sigma^{(n)}_{\mathrm{G}}$ in Fig.~\ref{fig:classes}.

\section{The color factor}\label{app:color}
The color factor in Eq.~\eqref{eq:Delta2},
\begin{equation*}
\Delta^{(f\rho f)}(n,j) = \big\langle f\big|\big(\mathcal{P}^{(1)}\circ\cdots\circ\mathcal{P}^{(j)}\circ\mathcal{C}\circ\Pi_\rho\circ\mathcal{P}^{(j+1)}\circ\cdots\circ\mathcal{P}^{(n)}\circ\mathcal{C}\big)[\boldsymbol{1}]\big|f\big\rangle\,, 
\end{equation*}
has $n$ emissions, the softer Coulomb exchange below all of them and the harder one between emissions $j$ and $j+1$.
Its closed form, Eq.~\eqref{eq:Deltaclosed}, is a product of two factors.
The $n-j$ emissions between the two exchanges give $(-2)^{n-j}$.
The $j$ emissions above both exchanges, together with the projection, give, for $\rho\neq f$ and $j\geq1$, \mbox{$(-1)^j(2^j-2)\dim R_\rho/N_c^5$}.
Between the exchanges, the color $\T_R$ on leg $a$'s side of the gap has been rotated.
The projection $\Pi_\rho[\boldsymbol{X}]=\boldsymbol{P}_\rho\boldsymbol{X}$ in Eq.~\eqref{eq:Delta2} keeps the part in the eigenspace of $\T_R^2$ with eigenvalue $\lambda_\rho$, where
\begin{equation}
\boldsymbol{P}_\rho=\prod_{\sigma\neq\rho}\frac{\T_R^2-\lambda_\sigma}{\lambda_\rho-\lambda_\sigma} \label{eq:Pdef}
\end{equation}
and $\sigma$ runs over the eigenvalues of $\T_R^2$.

Each $\mathcal{P}^{(l)}=\mathcal{P}_a^{(l)}+\mathcal{P}_b^{(l)}$ sums the emission of gluon $l$ off the two legs, Eq.~\eqref{eq:pair}.
Consider the action of $\mathcal{P}_i^{(l)}$, the emission off leg $i$, on an operator $\boldsymbol{X}$.
If $\boldsymbol{X}$ does not carry the color of gluon $l$, then $\boldsymbol{X}_{\mathrm{r}}=\boldsymbol{X}_{\mathrm{v}}=\boldsymbol{X}$, and the real term sums over the gluon's color, $\T_i^{(l)\dagger}\boldsymbol{X}\,\T_i^{(l)}=\T_i^c\boldsymbol{X}\,\T_i^c$.
The pair~\eqref{eq:pair} becomes a double commutator,
\begin{equation}
\mathcal{P}_i^{(l)}[\boldsymbol{X}] = \T_i^c\boldsymbol{X}\,\T_i^c-\tfrac12\big(\T_i^2\boldsymbol{X}+\boldsymbol{X}\T_i^2\big) = -\tfrac12\big[\T_i^c,[\T_i^c,\boldsymbol{X}]\big]\equiv\mathcal{K}_i[\boldsymbol{X}]\,. \label{eq:Kmap}
\end{equation}
Real and virtual cancel when $\boldsymbol{X}$ commutes with $\T_i$.
In general, the double commutator is the quadratic Casimir of the color rotations generated by $\T_i$, acting on $\boldsymbol{X}$ by commutation.
It takes a fixed value on each part of $\boldsymbol{X}$ that transforms as an irreducible representation.
Since leg $i$ is a quark or antiquark, $\boldsymbol{X}$ transforms under these rotations as $\mathrm{fund}\otimes\overline{\mathrm{fund}}=\mathbf{1}\oplus\mathrm{adj}$.
The Casimir vanishes on the singlet part $\boldsymbol{X}_{\mathbf{1}}$ and equals 2 on the adjoint part, in our units.
Hence
\begin{equation}
\mathcal{K}_i^{m}[\boldsymbol{X}]=(-1)^m\,\boldsymbol{X}+h(m)\,\boldsymbol{X}_{\mathbf{1}}\,,\qquad h(m)=\delta_{m0}-(-1)^m\,, \label{eq:Kpower}
\end{equation}
with $\mathcal{K}_i^0$ the identity.
The reason for separating the singlet part in this way will become clear below.

The $n-j$ emissions between the two exchanges act on $\mathcal{C}[\boldsymbol{1}]=\T_a^d\T_b^d$, which carries no gluon color.
On each leg, it is a single charge, with no singlet part, so by Eq.~\eqref{eq:Kpower} $\mathcal{P}_a^{(l)}$ and $\mathcal{P}_b^{(l)}$ each give $-1$.
Summing over the legs, each emission gives an overall factor of $-2$. Repeating this for the $n-j$ emissions gives
\begin{gather}
\big(\mathcal{P}^{(j+1)}\circ\cdots\circ\mathcal{P}^{(n)}\circ\mathcal{C}\big)[\boldsymbol{1}]=(-2)^{n-j}\,\mathcal{C}[\boldsymbol{1}]\,, \nonumber\\
\Delta^{(f\rho f)}(n,j)=(-2)^{n-j}\,\Delta^{(f\rho f)}(j,j)\,, \label{eq:enclosed}
\end{gather}
where $\Delta^{(f\rho f)}(j,j)$ has no emissions between the exchanges.

The charges of $b$ commute with $\T_a$ and $\T_R^2$, so we collect them on the right,
\begin{equation}
\Delta^{(f\rho f)}(j,j)=\big\langle f\big|\big(\mathcal{P}^{(1)}\circ\cdots\circ\mathcal{P}^{(j)}\big)\big[\T_a^d\,\boldsymbol{P}_\rho\,\T_a^c\;\T_b^d\T_b^c\big]\big|f\big\rangle\,. \label{eq:Xsplit}
\end{equation}
We start with the terms $\mathcal{P}_b^{(l)}$.
A gluon emitted off $b$ is not part of $\T_R$, so the operator in Eq.~\eqref{eq:Xsplit} does not carry its color, and Eq.~\eqref{eq:Kpower} applies to the charges of $b$.
For $m$ emissions off $b$,
\begin{equation}
\big(\mathcal{P}_b^{(1)}\circ\cdots\circ\mathcal{P}_b^{(m)}\big)\big[\T_a^d\,\boldsymbol{P}_\rho\,\T_a^c\;\T_b^d\T_b^c\big]=\T_a^d\,\boldsymbol{P}_\rho\,\T_a^c\;\mathcal{K}_b^{m}\big[\T_b^d\T_b^c\big]\,. \label{eq:Pb}
\end{equation}
For the terms $\mathcal{P}_a^{(l)}$, the gluon is part of $\T_R$, and Eq.~\eqref{eq:Kmap} does not apply directly.
Instead, we move $\boldsymbol{P}_\rho$ so that it acts on quark $a$ and the Coulomb gluon alone.
This is possible because each emission conserves the color on leg $a$'s side of the gap.

\begin{figure}[h]
\centering
\includegraphics[width=0.8\textwidth]{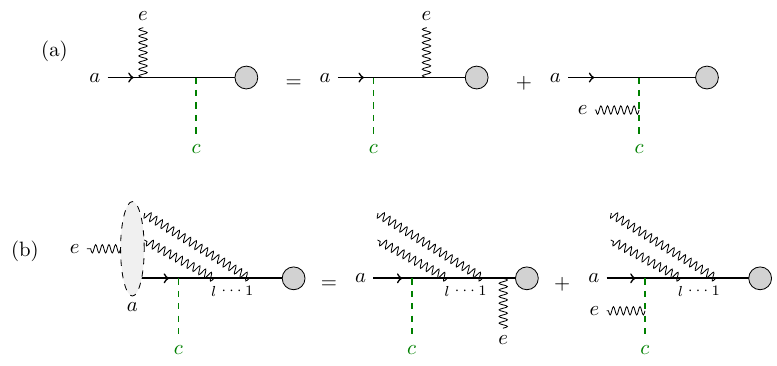}
\caption{(a) Eq.~\eqref{eq:Jdef}: a charge $e$ acting on quark $a$ after the Coulomb gluon $c$ equals the charge acting on the quark before the Coulomb plus the charge acting on the Coulomb gluon. (b) Eq.~\eqref{eq:Jcharge}: the total charge of $a$ and its real gluons $1,\dots,l$ (shaded), acting after the emissions and the Coulomb, equals the charge acting on the quark at the hard process plus the charge acting on the Coulomb gluon. The blob is the hard process.}
\label{fig:swap}
\end{figure}

Moving a charge $\T_a^e$ through the Coulomb exchange $\T_a^c$ gives
\begin{equation}
\T_a^e\,\T_a^c=\T_a^c\,\T_a^e+(\boldsymbol{t}^e)^{c'c}\,\T_a^{c'}=\T_a^{c'}\,(\boldsymbol{J}^e)^{c'c}\,,\qquad (\boldsymbol{J}^e)^{c'c}=(\boldsymbol{t}^e)^{c'c}+\T_a^e \delta^{c'c}\,, \label{eq:Jdef}
\end{equation}
with $\boldsymbol{t}$ the adjoint generators acting on the Coulomb gluon's index, $(\boldsymbol{t}^e)^{c'c} = i\sqrt{2/N_c} f^{ecc'}$.
After the Coulomb exchange, the charge $\T_a^e$ acts on the quark alone, and moved through it, it acts on both the quark and the Coulomb gluon, as $(\boldsymbol{J}^e)^{c'c}$ in Eq.~\eqref{eq:Jdef}, see Fig.~\ref{fig:swap}(a).
So, $\boldsymbol{J}$ is the total charge of quark $a$ and the Coulomb gluon.

With real emissions, let $\T_{R_l}$ be the charge of $a$ and its first $l$ real gluons, labelled in the order they act on $|f\rangle$, with $\T_{R_0}=\T_a$.
Color conservation at each emission gives
\begin{equation}
\T_{R_l}^e\,\T_a^{(l)}=\T_a^{(l)}\,\T_{R_{l-1}}^e\,. \label{eq:coherent}
\end{equation}
This is the identity of Fig.~\ref{fig:swap}(a) with $c$ replaced by the real gluon $l$, whose charge is part of $\T_{R_l}$.

\begin{figure}[h]
\centering
\includegraphics[width=0.62\textwidth]{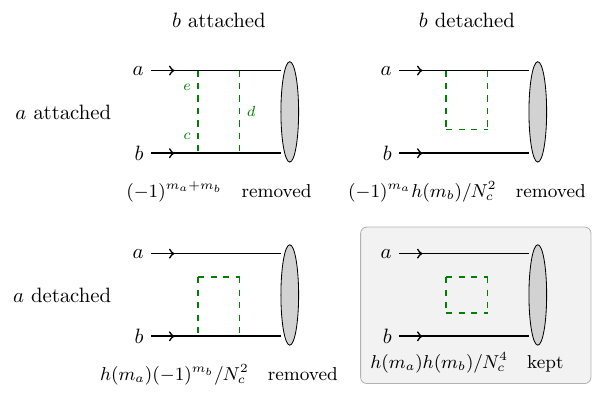}
\caption{The four terms of Eq.~\eqref{eq:wrho} after inserting Eq.~\eqref{eq:legfactor} on both legs, with their prefactors. On each leg the harder ($d$) and softer Coulomb gluons are attached to the leg or detached from it and contracted with each other. The softer one carries the index $e$ on $a$ and $c$ on $b$, connected by $(\boldsymbol{\Omega}_\rho)^{ec}$. Every term with an attachment is removed by Eq.~\eqref{eq:zerolaw}; only the closed loop (shaded) survives.}
\label{fig:attach}
\end{figure}

Let $\boldsymbol{v}_l^c=\T_a^c\,\T_a^{(l)}\cdots\T_a^{(1)}$ be $l$ real emissions off $a$ followed by the Coulomb charge, and $\T_R=\T_{R_l}$.
The gluon charges commute with $\T_a^c$, so Eqs.~\eqref{eq:Jdef} and~\eqref{eq:coherent} combine into
\begin{equation}
\T_R^e\,\boldsymbol{v}_l^c=\boldsymbol{v}_l^{c'}\,(\boldsymbol{J}^e)^{c'c}\,,\qquad \T_R^2\,\boldsymbol{v}_l^c=\boldsymbol{v}_l^{c'}\,(\boldsymbol{J}^2)^{c'c}\,. \label{eq:Jcharge}
\end{equation}
The real gluons drop out, Fig.~\ref{fig:swap}(b).
Projecting onto $\T_R^2=\lambda_\rho$ after $\boldsymbol{v}_l$ is therefore the same as projecting onto $\boldsymbol{J}^2=\lambda_\rho$ before it,
\begin{equation}
\boldsymbol{P}_\rho\,\boldsymbol{v}_l^c=\boldsymbol{v}_l^{e}\,(\boldsymbol{\Omega}_\rho)^{ec}\,, \label{eq:Ptransfer}
\end{equation}
with $\boldsymbol{\Omega}_\rho$ the projector onto the eigenvalue $\lambda_\rho$ of $\boldsymbol{J}^2$, Eq.~\eqref{eq:Pdef} with $\T_R^2\to\boldsymbol{J}^2$ and $\sigma$ over the eigenvalues of $\boldsymbol{J}^2$.
Since $\boldsymbol{J}$ is the charge of $\mathrm{fund}\otimes\mathrm{adj}$, $\rho$ runs over the channels $f,\pm$ of the main text.
It follows that, for $m$ emissions off $a$,
\begin{equation}
\big(\mathcal{P}_a^{(1)}\circ\cdots\circ\mathcal{P}_a^{(m)}\big)\big[\T_a^d\,\boldsymbol{P}_\rho\,\T_a^c\;\T_b^d\T_b^c\big]=\mathcal{K}_a^{m}\big[\T_a^d\T_a^e\big]\,(\boldsymbol{\Omega}_\rho)^{ec}\;\T_b^d\T_b^c\,. \label{eq:Pa}
\end{equation}
Each of the $j$ emissions in Eq.~\eqref{eq:Xsplit} is off $a$ or off $b$, and charges on different legs commute.
Inserting Eqs.~\eqref{eq:Pb} and~\eqref{eq:Pa} into Eq.~\eqref{eq:Xsplit}, and summing over the different assignments of the $j$ emissions to $a$ and $b$, gives
\begin{equation}
\Delta^{(f\rho f)}(j,j)=\sum_{m=0}^{j}\binom{j}{m}\,w_\rho(m,j-m)\,, \label{eq:binom}
\end{equation}
where $m$ is the number of emissions off $a$, and
\begin{equation}
w_\rho(m_a,m_b)=\langle f|\,\mathcal{K}_a^{m_a}[\T_a^d\T_a^e]\,(\boldsymbol{\Omega}_\rho)^{ec}\,\mathcal{K}_b^{m_b}[\T_b^d\T_b^c]\,|f\rangle\,. \label{eq:wrho}
\end{equation}
The singlet part of $\T_a^d\T_a^e$ is $\delta^{de}/N_c^2$ (in our units), so by Eq.~\eqref{eq:Kpower},
\begin{equation}
\mathcal{K}_a^{m}\big[\T_a^d\T_a^e\big]=(-1)^m\,\T_a^d\T_a^e+h(m)\,\frac{\delta^{de}}{N_c^2}\,, \label{eq:legfactor}
\end{equation}
and likewise for $b$.
Inserted into Eq.~\eqref{eq:wrho}, this gives the four terms of Fig.~\ref{fig:attach}: on each leg the two Coulomb gluons are either attached to the leg, the first term, or detached from it and contracted with each other, the singlet term.
Since $\Phi_f=0$ (cf.~Eq.~\eqref{eq:Phidef}), the channel $\rho=f$ does not enter Eq.~\eqref{eq:masterreorg}, and we need only $\rho=\pm$.
For these, every term with attached Coulomb gluons vanishes.
To see this, note that $\T_a^e$ contracts the softer Coulomb gluon into quark $a$, leaving a single quark, so it sees only the fundamental part of the pair: by Eq.~\eqref{eq:Jcharge} at $l=0$, $\T_a^e(\boldsymbol{J}^2)^{ec}=\lambda_f\T_a^c$.
Since $\boldsymbol{\Omega}_\pm$ contains the factor $\boldsymbol{J}^2-\lambda_f$, see Eq.~\eqref{eq:Pdef}, this gives, together with its Hermitian conjugate,
\begin{equation}
\T_a^e\,(\boldsymbol{\Omega}_\pm)^{ec}=0=(\boldsymbol{\Omega}_\pm)^{ec}\,\T_a^c\,. \label{eq:zerolaw}
\end{equation}
The first identity removes the two terms with the Coulomb gluons attached to $a$.
In the term with them detached on $a$ and attached to $b$, color conservation in the Born process, $\T_b^c|f\rangle=-\T_a^c|f\rangle$, gives $\T_b^d\T_b^c|f\rangle=-\T_a^c\,\T_b^d|f\rangle$, since charges on different legs commute.
This places $\T_a^c$ directly after $(\boldsymbol{\Omega}_\pm)^{ec}$, and the second identity in Eq.~\eqref{eq:zerolaw} removes this term.
Only the term with the Coulomb gluons detached on both legs survives, which is why we separated the singlet part in Eq.~\eqref{eq:Kpower}.
It is $h(m_a)\,h(m_b)\,N_c^{-4}\,\langle f|(\boldsymbol{\Omega}_\rho)^{cc}|f\rangle$.
In the Born state, the antiquark carries the anticolor $\bar\imath$ of the quark color $i$, so, normalized as in the main text ($\langle f|f\rangle=1$), $|f\rangle=\sum_i|i,\bar\imath\rangle/\sqrt{N_c}$.
As nothing acts on $b$, $\langle\bar\imath|\bar k\rangle=\delta_{ik}$ sets $k=i$, and
\begin{equation*}
\langle f|(\boldsymbol{\Omega}_\rho)^{cc}|f\rangle=\frac{1}{N_c}\sum_{i,k}\langle i,\bar\imath|(\boldsymbol{\Omega}_\rho)^{cc}|k,\bar k\rangle=\frac{1}{N_c}\sum_{i}\langle i,\bar\imath|(\boldsymbol{\Omega}_\rho)^{cc}|i,\bar\imath\rangle=\frac{\mathrm{Tr}\,\boldsymbol{\Omega}_\rho}{N_c}\,.
\end{equation*}
The sum runs over the colors $i$ of quark $a$ and $c$ of the Coulomb gluon, so it is the trace of $\boldsymbol{\Omega}_\rho$ over $\mathrm{fund}\otimes\mathrm{adj}$.
The projector $\boldsymbol{\Omega}_\rho$ has eigenvalue 1 on the $\dim R_\rho$ states of $R_\rho$ and 0 on all others, so $\mathrm{Tr}\,\boldsymbol{\Omega}_\rho=\dim R_\rho$. Hence
\begin{equation}
w_\rho(m_a,m_b)=h(m_a)\,h(m_b)\,\frac{\dim R_\rho}{N_c^5}\,,\qquad \rho\neq f\,. \label{eq:wclosed}
\end{equation}
Inserting Eq.~\eqref{eq:wclosed} into Eq.~\eqref{eq:binom} and summing over $m$ gives, with Eq.~\eqref{eq:enclosed}, Eq.~\eqref{eq:Deltaclosed} for $j\geq1$. At $j=0$, Eq.~\eqref{eq:binom} reduces to $w_\rho(0,0)\propto h(0)^2=0$, which is why Eq.~\eqref{eq:masterreorg} starts at $j=1$.

\end{document}